# Possible Origin of a Newly Discovered GeV Gamma-Ray Source Fermi J1242.5+3236

Xiu-Rong Mo[1], Ming-Hong Luo[1], Hong-Bin Tan[2], Qing-Wen Tang[1] , and Ruo-Yu Liu[2]
[1] Department of Physics, School of Physics and Materials Science, Nanchang University, Nanchang 330031, China; qwtang@ncu.edu.cn
[2] School of Astronomy and Space Science, Nanjing University, Nanjing 210023, China; ryliu@nju.edu.cn

## Abstract

Based on the first 13.4 yr of Fermi science data in the energy range from 300 MeV to 500 GeV, we discovered a bright GeV gamma-ray source with a ∼5.64$\sigma$ detection, named Fermi J1242.5+3236, which has an offset of about 0°.0996 from a nearby star-forming galaxy NGC 4631. When using the 12 yr' data, the detection significance of Fermi J1242.5+3236 is about 4.72$\sigma$. Fermi J1242.5+3236 is a steady point source without significant temporal variability and has a hard gamma-ray photon index of about $-1.60 \pm 0.24$. The spatial offset and the hard gamma-ray spectrum disfavor this source as the diffuse gamma-ray emission from this galaxy. This new source might have a possible origin of an unidentified background blazar, which is more likely a high-synchrotron-peaked blazar for its hard gamma-ray photon index. A follow-up optical observation would help distinguish origin of Fermi J1242.5+3236.

## 1. Introduction

In the gamma-ray sky above 100 MeV measured by the Fermi-Large Area Telescope (Fermi-LAT), the most prominent structure is the bright and diffuse band coincident with the Galactic plane. The component, which is called Galactic diffuse emission, is dominated by the pionic emission of cosmic rays (CRs) via interactions with the interstellar medium in the Milky Way. Neutral pions are produced in the interactions (schematically written as $p + p \to \pi^0$ + other products), which decay into gamma-rays ($\pi^0 \to \gamma + \gamma$). Besides the Galactic diffuse emission, a great many of Galactic and extragalactic gamma-ray sources above 100 MeV are also detected by Fermi-LAT. Most detected Galactic sources can be classified to pulsars (including millisecond pulsars, MSPs), pulsar wind nebulae (PWNe), supernova remnants (SNRs), and binaries. For extragalactic sources, blazars constitute the majority while a minority of them belongs to star-forming galaxies (SFGs) (Abdollahi et al. 2020, 2022).

SFGs are guaranteed gamma-ray emitters. Similar to the Galactic diffuse emission of the Milky Way, high-energy hadronic cosmic rays (CRs) generated in these SFGs can interact with interstellar medium therein as well and produce gamma-ray emission above dozens of MeV (Ginzburg & Syrovatskii 1964; Acero et al. 2009; VERITAS Collaboration et al. 2009; Abdo et al. 2010; Ackermann et al. 2012; Tang et al. 2014; Peng et al. 2016). However, detection of such type of galaxies is still rare, with respect to their large number in the universe. Including five local star-forming galaxies, there are only 12 star-forming galaxies detected in the high-energy gamma-ray bands (Ackermann et al. 2012; Tang et al. 2014; Peng et al. 2016; Xi et al. 2020). This is mostly due to their relatively low gamma-ray emissivity.

Nearby luminous-far-infrared star-forming galaxies are suggested to be hopeful gamma-ray sources (Ackermann et al. 2012). In this paper, we report a systematic search for possible $\gamma$−ray emission from galaxies in the IRAS Revised Bright Galaxies Sample, using 13.4 yr of $\gamma$−ray data collected by the Fermi-Large Area Telescope. Except for a nearby SFG, i.e., NGC 4631, we did not find any high-energy gamma-ray emission from the directions of these nearby galaxies.

NGC 4631, also known as the Whale galaxy, is an interacting galaxy located at 8.1 Mpc away. It exhibits one of the largest gaseous halos among observed edge-on galaxies, which may be undergoing considerable star formation activity and has a high star formation efficiency (Mora & Krause 2013; Mora-Partiarroyo et al. 2019). Its star formation is not dominated by the galactic nuclear region but extends over a large fraction of the galactic disk (Wang et al. 2001). Therefore, we first present here a new detection of a hard gamma-ray source in the northeast region of NGC 4631, and confirmed that there is no associated entry (within 1°.13) in two catalogs, such as the 10 yr point source catalogs (4FGL-DR2) and the 12 yr point source catalogs (4FGL-DR3) (Abdollahi et al. 2020, 2022). The rest of the paper is organized as follows. In Section 2, data analysis is presented. In Section 3, the physical origins of the new source are studied. Summary and conclusions are presented in Section 4.

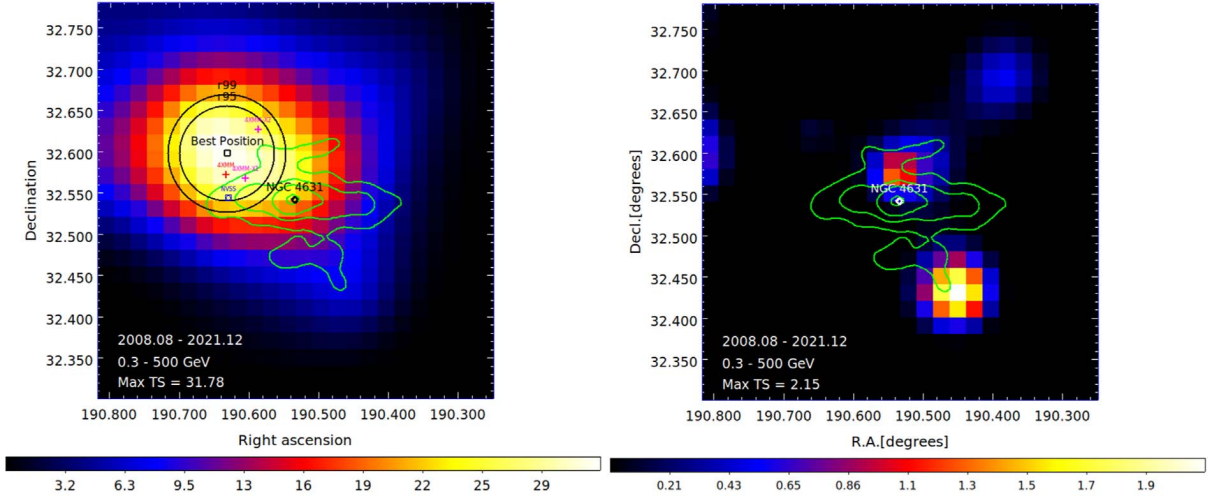

**Figure 1.** TS maps with bin size = $0°.02$ in the energy band 0.3–500 GeV centering at the galaxy NGC 4631. Left panel: TS map excluding Fermi J1242.5+3236, the central black empty diamond is the FIR center of NGC 4631, the red/pink crosses labeled as 4XMM/4XMM-X1/4XMM-X2 are the observations by the XMM-Newton survey and the blue boxsquare labeled as NVSS is the observation by the NRAO VLA Sky Survey. The northeast black empty box is the best localization for Fermi J1242.5+3236, around which two black circles are the 95% and 99% containment errors. Right panel: TS map including Fermi J1242.5+3236, the white empty diamond represents the FIR center of NGC 4631. In both panels, the green contours represent the FIR observations at 60 cm wavelength around NGC 4631 by the Infrared Astronomical Satellite (IRAS), with flux intensity of 250 Jy, 50 Jy, 10 Jy, 1 Jy from innermost layer to outermost layer respectively (Surace et al. 2004).

## 2. Data Analysis

### 2.1. Source Detection

We employed the Fermipy package to analyze the Fermi-LAT data, which is a python package that facilitates analysis of the Fermi-LAT data with the latest Fermi Science Tools (Wood et al. 2017). We downloaded the high-energy observation data from the Fermi-LAT data Server at the Fermi Science Support Center (FSSC).[3]

Both FRONT-type and BACK-type Pass 8 Source class events are used from 2008 August to 2021 December, whose region of interest (ROI) are selected within $15°$ of the far-infrared (FIR) center of NGC 4631, i.e., R.A. = $190°.5334$, decl. = $32°.5415$ (Surace et al. 2004). Excluding the events with a zenith angle larger than $90°$ is performed in order to remove the contaminant from the Earth limb.

To evaluate the detection significance of new gamma-ray sources, we performed the binned likelihood analyses. The background model comprises a diffuse-source model and a point-source model. The diffuse-source background model includes an isotropic extragalactic emission ("iso_P8R3_SOURCE_V_v1.txt") and a galactic diffuse emission ("gll_iem_v07.fits"), both of which are allowed to be free in the data analysis of source detection. The point-source background model is derived from the LAT 10 yr Source Catalog (4FGL-DR2), which comprises all public point sources within $25°$ of the ROI center.

---
[3] https://fermi.gsfc.nasa.gov/ssc/data/access/

Spectral parameters of the point sources within $6°.5$ region of the ROI center are allowed to be free.

Centering at FIR center of NGC 4631, we built a $0°.48 \times 0°.48$ map of the test statistic (TS), defined as TS $= -2(\ln L0 - \ln L)$, where $L_0$ is the maximum-likelihood value for null hypothesis and $L$ is the maximum-likelihood with the additional point source with a power-law spectrum. The criterion of a significance detection for a new gamma-ray source above the background is TS > 25 ($\sim 5\sigma$). As seen in the left panel of Figure 1, a significant gamma-ray excess with maximum TS of 31.78, corresponding to a $\sim 5.64\sigma$ detection, is found at the best-fit gamma-ray position of R.A. = $190°.6308$ and decl. = $32°.5976$, which is named as Fermi J1242.5+3236 according to its celestial coordinates. We used the Fermi-tool *gtfindsrc* to derive this best-fit position and calculate the error radius at 95% and 99% confidential levels, i.e., $r_{95} = 0°.0578$ and $r_{99} = 0°.0716$ respectively. The center of Fermi J1242.5+3236 is $0°.0996$ away from the center of NGC 4631. As seen in the TS map, the northeast region of NGC 4631 is overlapped with $r_{95}$ region of Fermi J1242.5+3236. Then we added a point source in the center of Fermi J1242.5+3236 and rebuilt the TS map. It is found that there is a little residual gamma-ray emission in the vicinity of NGC 4631, i.e., with the maximum TS value about 2.15, which can be found in the right panel of Figure 1. When using the 12 yr' data and background point source catalog of 4FGL-DR3, we derived the detection significance of Fermi J1242.5+3236 is about $4.72\sigma$, whose TS value is about 22.31.

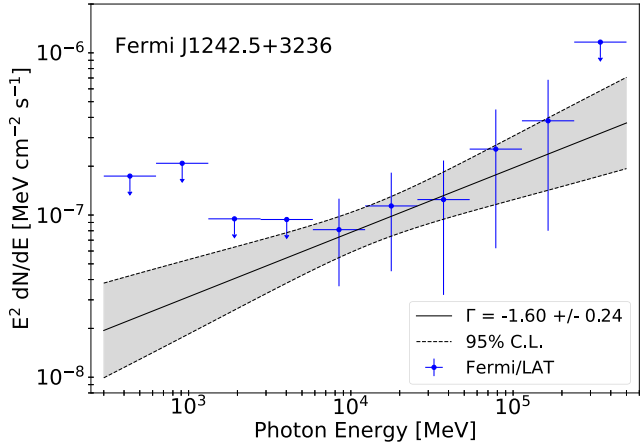

**Figure 2.** Spectral energy distribution of Fermi J1242.5+3236. Data points are plotted with 95% errors when TS > 4 while the upper limits are plotted at the 95% confidence levels when TS < 4. Solid black lines and gray shadows represent the best fit and the corresponding 95% confidence-level statistical uncertainty respectively.

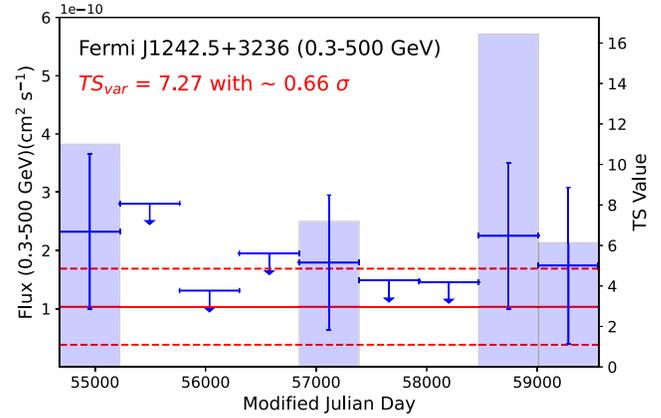

**Figure 3.** Lightcurve of Fermi J1242.5+3236 with nine time bins in the 0.3–500 GeV energy band. Data points are plotted with 95% errors when TS > 4 while the upper limits are plotted at the 95% confidence levels when TS < 4. The blue shadows are the TS values when TS > 4. The solid red line and dotted red line represent the best fit for the photon flux and the corresponding 95% confidential levels in the whole observation period, combining these nine time bins.

Replacing this point source with other extended sources, such as the Gaussian-distributed disks and the uniform disks in different radii, no significant improvement are found, which proves that Fermi J1242.5+3236 is actually a point source. Employing the spatial template of a uniform disk, we derived the best extended profile with $0°.041 \pm 0°.025$, whose TS value is only 0.84 larger than that as a point source.

### 2.2. Spectral Analysis of Fermi J1242.5+3236

As a common practise, the new point source can be described by a power-law function:

$$N(E) = K \left( \frac{E}{100 \text{ MeV}} \right)^{\Gamma} \quad (1)$$

where $K$ is the normalization, and $\Gamma$ is the photon index. For gamma-ray spectral energy distribution (SED) of Fermi J1242.5+3236, it is best-fitted with $\Gamma$ of $-1.60 \pm 0.24$, as shown in Figure 2. Here, gamma-ray photons with the energy range of 0.3–500 GeV are divided into 20 logarithmic bins for data analysis, which is plotted with 10 logarithmic bins by combining nearby two bins. Comparing with the typical photon index of Fermi-LAT point sources, the photon index of Fermi J1242.5+3236 is very hard. We also note that the individual TS values of spectral data points below about 5 GeV are all less than 4. We thus searched for possible associations in the third catalog of hard Fermi-LAT sources (3FHL), which is a catalog of sources detected above 10 GeV by the Fermi-LAT in the first 7 yr of data using the Pass 8 event-level analysis (Ajello et al. 2017). It is found that no 3FHL source is associated with Fermi J1242.5+3236 within $r_{95}$ region of Fermi J1242.5+3236.

### 2.3. Pulsation Analysis of Fermi J1242.5+3236

To investigate the periodic behaviors of Fermi J1242.5+3236, we searched for pulsations with the powspec task in the HEASoft package (version 6.28) by analyzing Fermi-LAT data. Three Fermi-LAT observations, such as period 1 from 2010 February to 2011 August, period 2 from 2014 September to 2016 March and period 3 from April 2019 to October 2020, are analyzed randomly.[4] An aperture with radius of 1° is selected. We used gtmktime tool to exclude events that are close to Earth's limb and that are within 5° of the Sun. The resultant photons are then processed by gtbary tool to apply barycentric corrections. We finally used the powspec to compute the Fourier power-law spectrum of these photons in three periods. No significant periodic signals are found.

### 2.4. Temporal Analysis of Fermi J1242.5+3236

We chosen an energy band of 0.3–500 GeV to generate its lightcurve (LC). The observation period of 13.4 yr was divided into nine linearly equal-time bins, and the data of each time bin was fitted by the similar likelihood analysis. The LC of Fermi J1242.5+3236 can be found in Figure 3, in which the upper-limit fluxes at 95% confidence level are given when TS < 4 in four time bins. We also plot the best-fit photon flux during whole 13.4 yr periods, such as $(1.03 \pm 0.65) \times 10^{-10}$ cm$^{-1}$ s$^{-1}$. It can be seen that Fermi J1242.5+3236 has no obvious variability, which can be found in Figure 3.

To examine whether Fermi J1242.5+3236 does have variability, the significance of variability by the variability index (TS$_{\rm var}$) is also tested. TS$_{\rm var}$ is commonly used to estimate the significance of variability for a source, which is defined as:

$$\mathrm{TS}_{\rm var} = 2\sum_{i=1}^{N}[\log(\mathcal{L}_i(F_i)) - \log(\mathcal{L}_i(F_{\rm glob}))] \qquad (2)$$

where $\mathcal{L}_i$ is the likelihood corresponding to bin $i$, $F_i$ is the best-fit flux for bin $i$, and $F_{\rm glob}$ is the best-fit flux for the full time, assuming a constant flux (Nolan et al. 2012; Gu et al. 2022). For nine bins, the critical value of TS$_{\rm var} \geqslant 20.09$ is used to identify variable sources at a 99% confidence level. The resultant value of TS$_{\rm var} = 7.27$ with $0.66\sigma$ for the LC of nine time bins suggests that there is no significant variability for the photon fluxes of Fermi J1242.5+3236. The similar results are found in the energy bands of 1–500 GeV and 5–500 GeV, which are plotted in Figure A1.

---

[4] https://heasarc.gsfc.nasa.gov/xanadu/xronos/examples/powspec.html/

## 3. Physical Origins of Fermi J1242.5+3236

### 3.1. Search for Possible Counterparts of Fermi J1242.5+3236

Since we did not find a periodic signal of Fermi J1242.5+3236 in Section 2.3, a pulsar origin or a binary origin is not favored. On the other hand, other common Galactic gamma-ray sources such as PWNe and SNRs are generally located as the Galactic disk, while this source is a at the Galactic latitude of $b = 84°.2$. The position is inconsistent with a Galactic origin, unless the source distance is quite close to us, within a few hundred parsecs. However, given such a close distance, a PWN or SNR would appear as an extended source while this source is a point-like source. Therefore, the Galactic origin of Fermi J1242.5+3236 is not favored.

Given the GeV flux of the source being $(6.67 \pm 2.45) \times 10^{-13}\,\mathrm{erg\,cm^{-2}\,s^{-1}}$, we can estimate its luminosity to be $(5.23 \pm 1.92) \times 10^{39}\,\mathrm{erg\,s^{-1}}$ assuming the emission is originated from NGC 4631. Previous measurements of Fermi-LAT on Local Group galaxies and nearby star-forming galaxies suggest an empirical correlation between the 0.1 and 100 GeV gamma-ray luminosity and the infrared luminosity of the galaxy (e.g., Ackermann et al. 2012; Xi et al. 2020), $\log(L_\gamma/\mathrm{erg\,s^{-1}}) = (1.17 \pm 0.07)\log(L_{8-1000\mu m}/10^{10}L_\odot) + (39.28 \pm 0.08)$. Given the infrared luminosity of this galaxy to be $2 \times 10^{10}L_\odot$, the gamma-ray luminosity derived from this empirical relation is about $4.29^{+5.12}_{-2.28} \times 10^{39}\,\mathrm{erg\,s^{-1}}$, which is consistent with the measured one. However, the position of the gamma-ray source deviates from the center of the galaxy by about $0°.1$. On the other hand, the measured gamma-ray spectrum is quite hard, i.e., $\Gamma = 1.6$, compared with that of the diffuse gamma-ray flux of the Milky Way and other nearby SFGs with $\Gamma > 2$. These two properties of the source disfavor a star-forming galaxy interpretation. Such a large luminosity also makes the source unlikely the astrophysical objects residing in the galaxy, such as gamma-ray binaries or pulsar wind nebula. This leads us to speculate Fermi J1242.5+3236 to be a distant gamma-ray blazar, which is of the largest population in the Fermi-LAT catalog. Comparing with the spectra of other blazars detected by Fermi-LAT, the hard spectrum of this source seems to be more consistent with a high-synchrotron-peaked blazar (HSP, Ajello et al. 2020, 2022).

Then we searched for the possible observations within $r_{95}$ region of Fermi J1242.5+3236 in low-energy-band surveys, such as the X-ray band and the radio band. In the X-ray bands, we found three X-rays sources in the XMM−Newton Serendipitous Source Catalog of dr11 (Webb et al. 2020), i.e., 4 XMM J124231.8+323419 (4XMM), 4 XMM J124224.9+323404 (4XMM-X1) and 4 XMM J124220.6+323737 (4XMM-X2). In the radio bands, NVSS J124231+323237 (NVSS) is found in NRAO VLA Sky Survey Catalog (Condon et al. 1998). The information of all these observations is presented in Table 1. Their fluxes are plotted in Figure A2.

We also checked possible low-energy counterparts of Fermi J1242.5+3236 within the Veron Catalogue of Quasars & AGN, 13th Edition (Véron-Cetty & Véron 2010), the 5th Edition of the Roma BZCAT Multi-frequency Catalogue of Blazars (Massaro et al. 2015), the WIBRALS − WISE Blazar−like Radio−Loud Source (WIBRaLS) Catalog (D'Abrusco et al. 2014), and the CRATES − CRATES Flat−Spectrum Radio Source Catalog (Healey et al. 2007). However, no candidate source is found within the $r_{95}$ region of Fermi J1242.5+3236 in these catalogs. As such, we will discuss from a theoretical point of view the possible origin of Fermi J1242.5+3236 in the scenarios of a star-forming galaxy origin and of a blazar origin, respectively, in the next section, in which we regard the flux of NVSS and 4XMM (which is the closest X-ray source to Fermi J1242.5+3236 among the three searched X-ray sources) as the upper limit of the flux for Fermi J1242.5+3236 at low energy bands.

### 3.2. Physical Model

#### 3.2.1. Hadronic Case in the SFG Scenario?

In the hadronic case, we modeled the high-energy observations of Fermi J1242.5+3236 by the pion-decay induced emission (PD) from the primary protons interacting with the interstellar medium in NGC 4631. We assumed the proton distribution to be a power-law formula with an exponential cutoff (PLEC), such as

$$N(E) = A\left(\frac{E}{E_{\rm piv}}\right)^{-\alpha_p}\exp\left(-\frac{E}{E_C}\right) \qquad (3)$$

where $A$ is the amplitude, $\alpha_p$ is the spectral index, $E$ is the proton energy, and $E_C$ is the cutoff energy of the proton spectrum, which is fixed at 30 TeV. To calculate the products of the $p-p$ collisions, we employ the model by Kamae et al. (2006). The distance of Fermi J1242.5+3236 was adopted to be 8.1 Mpc, same as the galaxy NGC 4631.

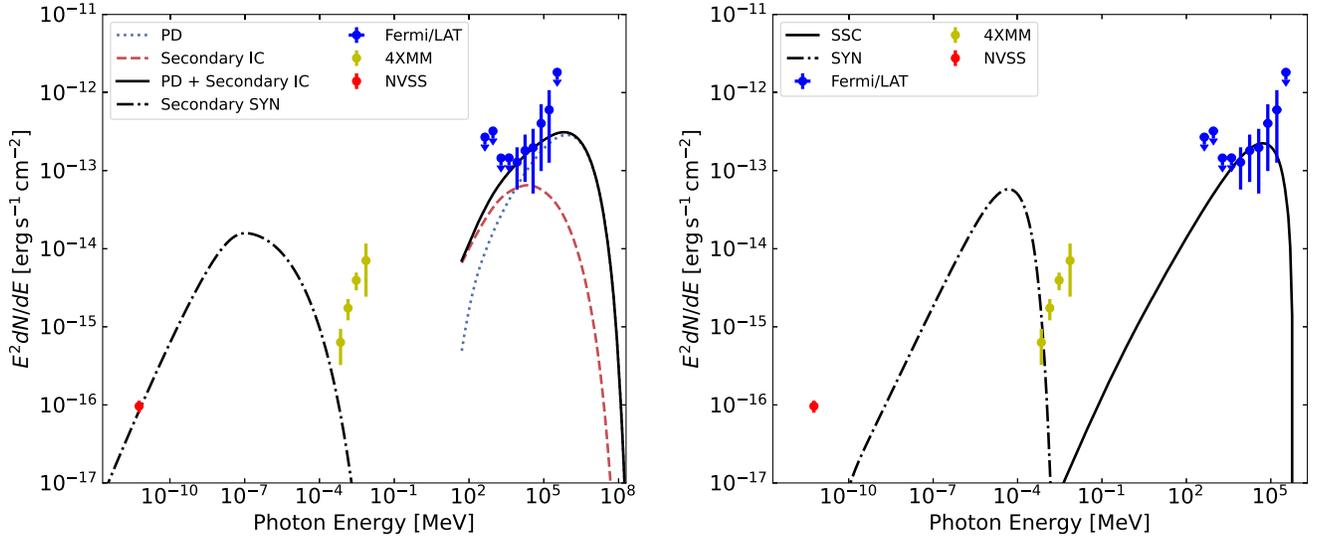

**Figure 4.** Left panel: Theoretical expectation of the hadronic-model SED in the SFG scenario. The blue dotted line is the pionic decay induced emission (PD) from the primary protons. The red dashed line is the inverse Compton emission (Secondary IC) from the secondary electrons while the dashed–dotted line is the synchrotron emission (Secondary SYN) from the same secondary electrons with the magnetic field $B = 1.8\ \mu G$. The spectral index of the proton spectrum is $\alpha_p = 1.45$. Right panel: Theoretical expectation of the leptonic-model SED in the blazar scenario. The black solid line is the synchrotron self-Compton radiation (SSC) of the primary electrons and the dashed–dotted line is the synchrotron emission (SYN) from the same electrons. The electron injection spectral index is $\alpha_e = 1.50$, injection luminosity is $L_{e,\mathrm{inj}} = 10^{39.5}\ \mathrm{erg\ s^{-1}}$, and the magnetic field is assumed to be $B = 0.05\ G$. The maximum energy of electrons is $\gamma_{\max} = 10^{5.15}$, and the minimum energy is $\gamma_{\min} = 10^2$. The radius of the radiation zone is $R = 10^{15}$ cm and the Doppler factor is $\delta_D = 10$. Gamma-ray data are derived from the LAT observation of Fermi J1242+323 in this work. Radio data are from the NRAO VLA Sky Survey while the X-ray data are from the XMM-Newton Survey.

Table 1
The Best-fit Position of Fermi J1242.5+3236 and the Nearby Observations by Other Surveys

| Name | Label | R.A. (degree) | Decl. (degree) | Error Radius (arcsecond) | Offset (arcsecond) |
|---|---|---|---|---|---|
| Fermi J1242.5+3236 | | 190.6308 | 32.5976 | 207.9176 | |
| 4XMM J124231.8+323419 | 4XMM | 190.6328 | 32.5722 | 1.35 | 91.5415 |
| 4XMM J124224.9+323404 | 4XMM-X1 | 190.6042 | 32.5678 | 1.11 | 134.0762 |
| 4XMM J124220.6+323737 | 4XMM-X2 | 190.5861 | 32.6270 | 1.73 | 171.9443 |
| NVSS J124231+323237 | NVSS | 190.6292 | 32.5438 | 3.34 | 193.6296 |

In addition to gamma-rays, secondary electrons and positrons are produced as well in the PD process, which results in an inverse Compton emission (Secondary IC) and a synchrotron emission (Secondary SYN) in the radiation field and the magnetic field of the galaxy respectively. Seed photons in the cosmic microwave background (CMB), the FIR radiation field and the near-infrared (NIR) radiation field are all considered in the secondary inverse Compton scattering processes. NGC 4631 has similar observations as the Milky Way in several IR surveys, such as the luminosity at 25 $\mu$m of about $7 \times 10^{22}\ \mathrm{W\ Hz^{-1}}$, the total IR luminosity between 8 and 1000 $\mu$m of about $2 \times 10^{10} L_\odot$. For the FIR and NIR radiation field, we therefore consider a greybody spectrum of a temperature of 30 and 5000 K respectively with the energy density $0.3\ \mathrm{eV/cm^3}$ for both radiation fields. We fit the data by the secondary IC and SYN components as well as the PD component. The best-fit model is shown in the left panel of Figure 4, in which $\alpha_p = 1.45$, $E_{\mathrm{piv}} = 2.5$ TeV are derived. In order not to overshoot the X-ray and radio data, an upper limit for the magnetic field $B = 1.8\ \mu G$ is needed. The resultant parameters are listed in Table 2. The very hard injection spectrum makes the star-forming galaxy interpretation quite dubious.

### 3.2.2. Leptonic Case in the Blazar Scenario?

Fermi J1242.5+3236 may be also explained as a background gamma-ray blazar. In this scenario, the main radiation processes considered are the synchrotron radiation and the synchrotron self-Compton (SSC) radiation. We assumed a PL type of the injection electron distribution $n_e$. Considering the cooling effect, the steady state electron density distribution in the blazar's radiation zone can be

$$N(E) = A\left(\frac{E}{E_{\mathrm{piv}}}\right)^{-\alpha_e} \quad (4)$$

where $\alpha_e$ is the injection spectral index and $E_{\mathrm{piv}} = 1$ TeV is the pivot energy. The amplitude $A = \frac{L_{e,\mathrm{inj}}}{V \int E_e n_e dE_e} \min(t_{\mathrm{cool}}, t_{\mathrm{esc}})$ is determined by the injection luminosity $L_{e,\mathrm{inj}}$, the volume of the radiation zone, which is assumed to be a spherical blob, i.e., $V = (4/3)\pi R^3$. $t_{\mathrm{esc}} = 10R/c$ is the escape timescale following the treatment of Böttcher et al. (2013) and Gao et al. (2017), and $t_{\mathrm{cool}} = 3m_e c^2/4\sigma_T \gamma (u_B + u_{\mathrm{syn}})$ is the radiative cooling timescale in the comoving frame of the radiation zone. $u_B$ and $u_{\mathrm{syn}}$ here are the energy density of the magnetic field and the synchrotron radiation field respectively. The smaller one of between $t_{\mathrm{esc}}$ and $t_{\mathrm{cool}}$ determines the amount of injected electrons that can be accumulated in the radiation zone.

Table 2
Best-fit Proton/electron Spectral Parameters for Fermi J1242.5+3236 in Two Scenarios

| Component | $\log_{10}(A)$ (GeV$^{-1}$) | $E_0$ (TeV) | $\alpha$ | $E_C$ (TeV) | $B$ | Candidate System |
|---|---|---|---|---|---|---|
| Hadronic case | | | | | | |
| PD (proton) | 51.0 | 2.5 | 1.45 | 30.0 (fixed) | ... | SFG |
| 2nd SYN | ... | ... | ... | ... | <1.8 $\mu$G | ... |
| Leptonic case | | | | | | |
| SSC (electron) | 42.7 | 1.0 | 1.50 | ... | ... | Blazar |
| SYN | ... | ... | ... | ... | <0.05 G | ... |

A simple one-zone quasi-stationary model is employed, in which the SSC radiation accounts for the gamma-ray flux, whereas the synchrotron radiation should not overproduce the radio and the X-ray emission. The distance of the blazar is unknown and we assume a redshift of $z = 0.1$. At such a redshift, the required electron injection luminosity is $L_{e,\mathrm{inj}} = 10^{39.5}$ erg s$^{-1}$. To explain the gamma-ray spectrum, electron injection spectral index need be around $\alpha_e = 1.5$. In order not to overproduce the low-energy flux, the magnetic field must be smaller than $B = 0.05$ G. The maximum energy of electrons is $\gamma_{\mathrm{max}} = 10^{5.15}$, and the minimum energy is $\gamma_{\mathrm{min}} = 10^2$. The radius of the radiation zone is $R = 10^{15}$ cm and the Doppler factor is $\delta_D = 10$. The upper-limit magnetic field is 0.05 G. The fitting result is shown in the right panel of Figure 4, and the resultant parameters are presented in Table 2. Note that the values of model parameters employed here are trivial, as the true redshift is unknown. For a higher redshift, we may increase the electron injection luminosity and the magnetic field to fit the gamma-ray flux. The main point is to show a blazar interpretation can work for this source. Assuming Fermi J1242.5+3236 being a blazar, we calculated the upper limit of gamma-ray luminosity of NGC 4631 with a common photon index of $-2.2$ above 100 MeV at 95% confidence level $L_{\gamma,\mathrm{UL95}} = 5.05 \times 10^{39}$ erg s$^{-1}$, which is slightly lower than that reported in Ackermann et al. (2012).

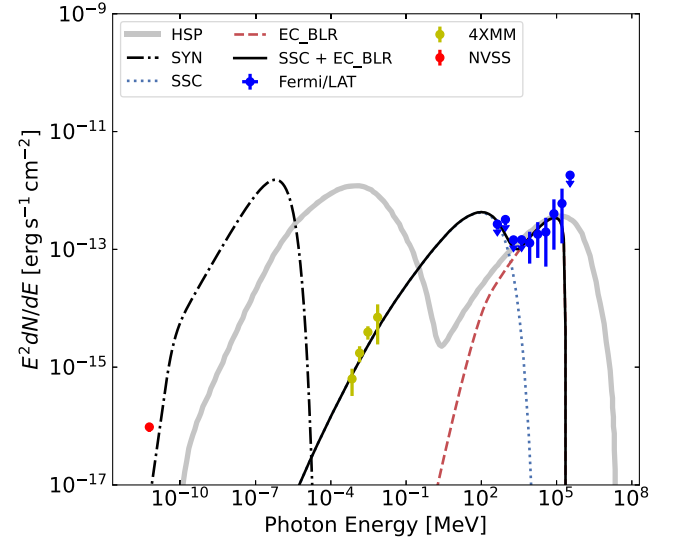

**Figure 5.** Same as the right panel in Figure 4, but assuming the observed X-rays (4XMM) originated from Fermi J1242.5+3236. The electron injection spectral index is $\alpha_e = 1.50$, injection luminosity is $L_{e,\mathrm{inj}} = 10^{39.95}$ erg s$^{-1}$, and the magnetic field is assumed to be $B = 0.01$ G. The maximum energy of electrons is $\gamma_{\mathrm{max}} = 10^{4.23}$, and the minimum energy is $\gamma_{\mathrm{min}} = 10^2$. The radius of the radiation zone is $R = 10^{16.18}$ cm and the Doppler factor is $\delta_D = 30$. The spectrum of the photons from BLR is a graybody spectrum peaking at $2 \times 10^{15}\delta_D$ Hz at the comoving frame, with energy density $u_{\mathrm{BLR}} = 10^{-4.70}$ erg cm$^{-3}$. The thick gray line is the typical SED of the high-synchrotron-peaked blazar (HSP, Ghisellini et al. 2017).

Alternatively, assuming 4XMM source originated from Fermi J1242.5+3236, we may model X-rays with the SSC radiation and gamma-rays with the external Compton (EC) scattering on the radiation of the broad line region (BLR), implying the blazar to be an FSRQ. The spectrum of the BLR photons is considered as graybody spectrum, which peaks at $2 \times 10^{15}\delta_D$ Hz at the comoving frame and the energy density is assumed as $u_{\mathrm{BLR}} = 10^{-4.70}$ erg cm$^{-3}$. The fitting result is shown in Figure 5, in which a typical SED of the Fermi HSP is plotted (Ghisellini et al. 2017). As seen in Figure 5, Fermi J1242.5 +3236 is consistent with the typical Fermi HSP in the GeV band, although they have a great difference in the X-ray band.

## 4. Summary and Conclusion

In this work, we reported discovery of a hard GeV source in the northeast region of the star-forming galaxy NGC 4631, named Fermi J1242.5+3236. It is proved being a steady point sources with none pulsation, around which several X-ray and radio observations are found. Using the X-ray flux and radio flux from the closest sources as an upper limit of this gamma-ray source, we found that the gamma-ray data may be reproduced by either the proton-proton collisions between cosmic rays in the galaxy and the interstellar medium or the inverse Compton radiation of electrons in the scenario of a background blazar. Although the inferred luminosity is consistent with that from the empirical relation between the infrared luminosity and gamma-ray luminosity of a star-forming galaxy, the former scenario is disfavored given the large systematic offset between the position of the gamma-ray source and the center of the galaxy, as well as the unusually hard cosmic ray spectrum compared with other star-forming galaxies. If Fermi J1242.5+3236 is truly a blazar, we would expect multiwavelength counterpart at lower energies. A deep follow-up optical or radio observation on the source would help to identify its origin.

## Acknowledgments

We are grateful to the referee for his/her helpful suggestions. We thank Xiang-Yu Wang, Dong Xu, Yun-Fei Xu, Ze-Rui Wang, Hai-Ming Zhang, Fang-Kun Peng and Shao-Qiang Xi for helpful discussions. This work made use of the High Energy Astrophysics Science Archive Research Center (HEASARC) Online Service at the NASA/Goddard Space Flight Center (GSFC). This work is supported by the Jiangxi Provincial Natural Science Foundation under grant 20212BAB201029 and the National Natural Science Foundation of China under grants 11903017, 11975116, 12065017, and U2031105.

## Appendix

Figure A1 shows the light curves of Fermi J1242.5+3235 in two gamma-ray bands, e.g., top panel for 1-500 GeV and bottom panel for 5-500 GeV. Figure A2 shows the nearby low-energy observations of Fermi J1242.5+3235, such as in X-ray bands and radio band.

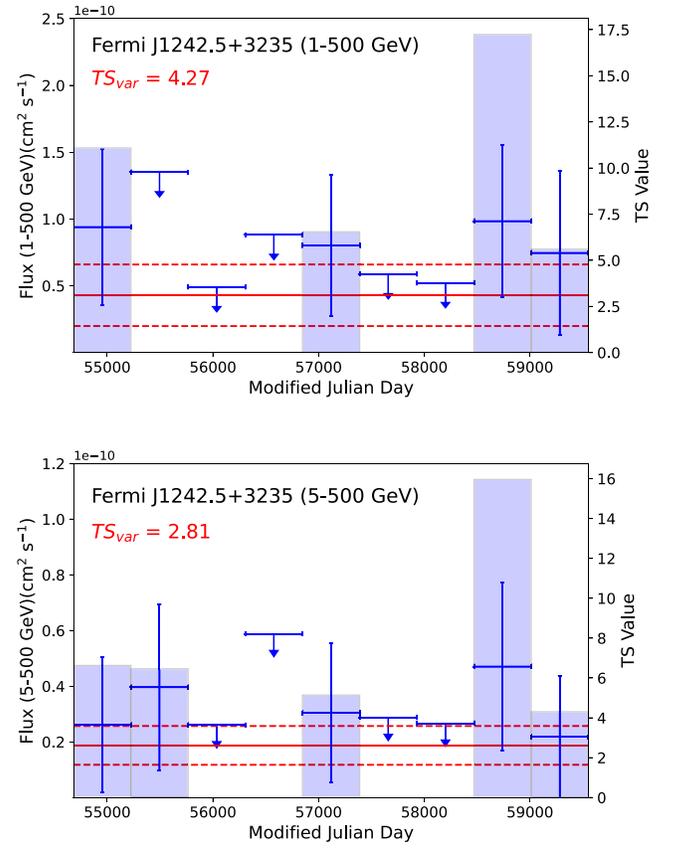

**Figure A1.** Same as Figure 3, but the top panel for 1–500 GeV and the bottom panel for 5–500 GeV.

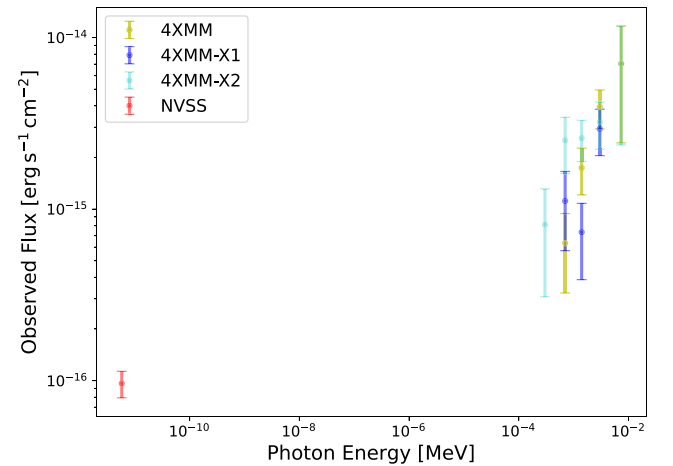

**Figure A2.** Nearby observations of Fermi J1242.5+3236 in the X-ray or radio bands.